    \documentclass[conference]{IEEEtran}
    \IEEEoverridecommandlockouts

    \makeatletter
    \def\markboth#1#2{\def\leftmark{\@IEEEcompsoconly{\sffamily}\MakeUppercase{\protect#1}}%
    \def\rightmark{\@IEEEcompsoconly{\sffamily}\MakeUppercase{\protect#2}}}
    \makeatother

    \usepackage[margin=0.7 in]{geometry}
    \usepackage{kantlipsum}
    \usepackage{setspace}

    \usepackage[english]{babel}
    \usepackage{xcolor}
    \usepackage{lipsum}
    \usepackage{caption}
    \usepackage{cite}
    \usepackage[pdftex]{graphicx}
    \usepackage{subcaption}
    \usepackage{amsmath}
    \usepackage{booktabs}
    \usepackage{colortbl}
    \usepackage{amsfonts}
    \usepackage{amssymb}
    \usepackage{amsthm}
    \usepackage{array}
    \usepackage{dblfloatfix} 

    \usepackage{verbatim}
    \usepackage{listings}
    \usepackage{algorithm}
    \usepackage{algpseudocode}
    \usepackage{url}
    \usepackage{enumerate}
    \usepackage{multirow}

    \definecolor{LightBlue}{rgb}{0.5,0.5,1}
    \definecolor{LightRed}{rgb}{1,0.5,0.5}
    \definecolor{LightYellow}{rgb}{1,0.85,0}

    \usepackage{epsfig}
    \usepackage{epstopdf}
    \usepackage{multicol}
    \usepackage[font=footnotesize]{caption}

    \usepackage{algorithm}

    \makeatletter
    \def\BState{\State\hskip-\ALG@thistlm}
    \makeatother

    \renewcommand{\arraystretch}{2}

    \newcommand{\C}{\mathbf{C}}
    \newcommand{\E}{\mathbf{E}}

    \newcommand{\bi}{\begin{itemize}}
    \newcommand{\ei}{\end{itemize}}
    \newcommand{\be}{\begin{equation}}
    \newcommand{\ee}{\end{equation}}

    \newtheorem{proposition}{Proposition}
    \newtheorem{remark}{Remark}

    \def\beq{\begin{equation}}
    \def\eeq{\end{equation}}
    \def\beqa{\begin{eqnarray}}
    \def\eeqa{\end{eqnarray}}
    \def\beqan{\begin{eqnarray*}}
    \def\eeqan{\end{eqnarray*}}

    \def\SIR{\mathsf{SIR}}

    \def\({\left(}
    \def\){\right)}
    \def\[{\left[}
    \def\]{\right]}

    \def\C{C_{\SIR}}

    \title{ Stochastic Geometric Coverage Analysis in mmWave Cellular Networks with a Realistic Channel Model}
    \author{{{\bf Mattia Rebato}$^*$, {\bf Jihong Park}$^\dagger$, {\bf Petar Popovski}$^\dagger$, {\bf Elisabeth De Carvalho}$^\dagger$, {\bf Michele Zorzi}$^*$
    }\\
    $^*$University of Padova, Italy \\
    $^\dagger$Aalborg University, Denmark\\
    \small{$\{$\texttt{rebatoma}, \texttt{zorzi}$\}$\texttt{@dei.unipd.it}, $\{$\texttt{jihong},  \texttt{petarp}, \texttt{edc}$\}$\texttt{@es.aau.dk} }
    }

\begin{document}
    \maketitle
    \begin{abstract}
     Millimeter-wave (mmWave) bands have been attracting growing attention as a possible candidate for next-generation cellular networks, since the available spectrum is orders of magnitude larger than in current cellular allocations.
    To precisely design mmWave systems, it is important to examine mmWave interference and $\SIR$ coverage under large-scale deployments.
    For this purpose, we apply an accurate mmWave channel model, derived from experiments, into an analytical framework based on stochastic geometry.
    In this way we obtain a closed-form $\SIR$ coverage probability in large-scale mmWave cellular networks.
    
    \end{abstract}
      \smallskip

    \section{Introduction}
    \label{introduction}
    A key enabler for the upcoming 5G cellular networks will be the use of millimeter-wave (mmWave) frequencies~\cite{mmwave3gpp}.
    The available mmWave spectrum above 6~GHz is expected to be 20-100 times larger than that at the sub 6~GHz frequencies currently used for cellular services.
    MmWave will thereby be able to cope with the relentless growth of cellular user demand.
    To enjoy the abundant bandwidth, it is necessary to sharpen transmit/receive beams in order to compensate for the significant distance attenuation of the desired mmWave signals~\cite{roh2014,mmwave3gpp}.
    The directionality of the mmWave transmissions can cause intermittent but strong interference to the neighboring receivers. 
    For example, sharpened transmit beams reduce the interfering probability from the main lobe, while increasing the signal strength of the main lobe.

    To design downlink 5G mmWave systems, it is therefore important to examine mmWave inter-cell interference and the corresponding signal-to-interference ratio ($\SIR$) coverage under large-scale deployments.
    Recent works \cite{bai14,bai15,direnzo2015,park2016} have investigated mmWave network coverage by using stochastic geometry, a mathematical tool that captures random interference behaviors in a large-scale network.
    For mathematical tractability, such approaches resort to assuming Rayleigh small-scale fading, i.e., an exponentially distributed fading power gain. This is not always realistic when modeling the sparse scattering characteristics of mmWave due to scarcity of multipath in the directed beam. 
    At the cost of compromising tractability, several works have detoured this problem by considering generalized fading distributions including Nakagami-m~\cite{direnzo2015,bai15} and log-normal~\cite{direnzo2013}.
    Nevertheless, such generic fading models have not been compared with real mmWave channel measurements, and may therefore over/under-estimate the exact fading behaviors.

In addition, existing works use a simplified beam pattern to model the beamforming gain.
A precise characterization of the impact of side lobes and backward propagation is critical when evaluating the beamforming gain which cannot be negligible even when the number of antennas is very large (especially for the interference gains).

\begin{figure}[t!]
    \centering
    \includegraphics[width=1\columnwidth]{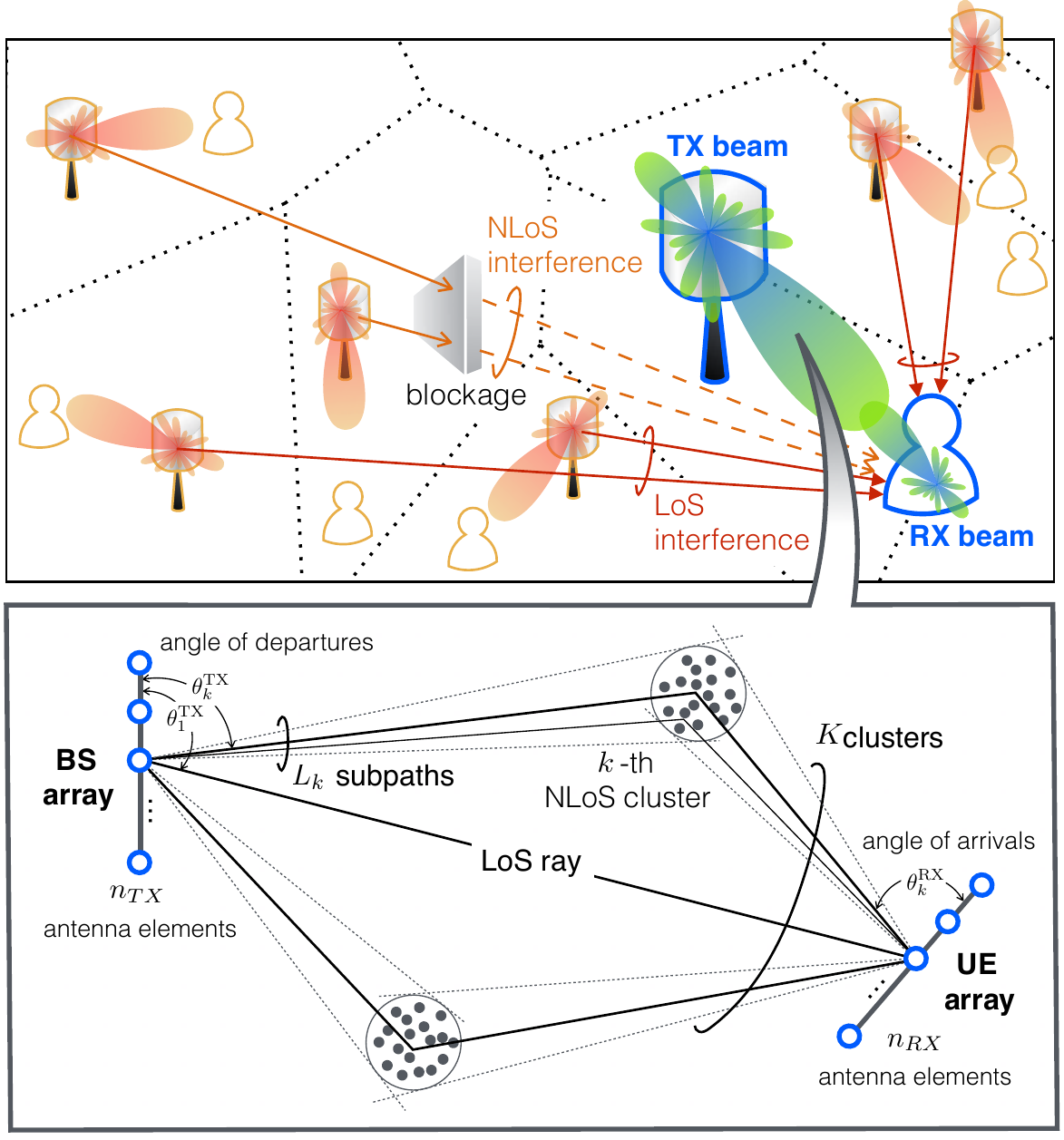}
    \caption{Illustration of our mmWave network (top) and the channel model of each link (bottom). Channel parameters follow from a measurement-based mmWave channel model provided by the NYU Wireless Group~\cite{akdeniz14}.}
    \label{globalview}
    \end{figure}
    Motivated by these preceding works and by the need to properly capture propagation behaviors for an accurate performance analysis, in this paper we aim at deriving downlink mmWave $\SIR$ coverage based on a realistic measurement-based channel model while still providing a tractable analysis.
    In addition to mmWave small-scale fading, as illustrated in Fig. 1 we also incorporate mmWave channel and transceiver characteristics such as antenna patterns and beamforming gains, directly following from a measurement-based mmWave channel model operating at 28~GHz provided by the New York University (NYU) Wireless Group explained in~\cite{akdeniz14,samimi15,mezzavilla15,ford16}, and adopted in our previous work~\cite{rebato16}.

    To this end, at a randomly picked (i.e., typical) user equipment (UE), we first define the \emph{aligned gain}, that indicates aggregate received power gain for the desired signal link at unit distance. This term captures the effects of small-scale fading and transmit/receive antenna patterns with directional beamforming. Similarly, we define the \emph{misaligned gain} for interfering links, which includes main and side lobes interfering patterns compared to the aligned gain. By curve fitting with a Monte Carlo simulation based on the NYU mmWave channel model, we derive the aligned and misaligned gain distributions. 
    
    Our analysis leads to the following two observations. First, the aligned gain is well-fitted by an \emph{exponential} distribution despite the fact that the scattering environment for a mmWave channel is sparse due to scarce multipath in the directed beam (see \textbf{Remark~1} in Section III). Second, the misaligned gain follows a \emph{log-logistic} distribution (\textbf{Remark~2}), which is lower and upper bounded, respectively, by a \emph{Burr} distribution and a \emph{log-normal} distribution (\textbf{Remark~3}). 
    We apply these results to an $\SIR$ coverage expression by using stochastic geometry (\textbf{Proposition~1}).
    This provides a way to tractably incorporate realistic mmWave channel characteristics into a stochastic geometric coverage analysis, thereby bridging the gap between analytical and simulation-based approaches.
    To the best of our knowledge, this is the first paper that incorporates a realistic mmWave channel model in a stochastic geometric analysis.

    \section{System Model}
    \label{system_model}
	Consider a downlink mmWave cellular network where  the locations of the base stations (BSs) are uniformly distributed with density $\lambda_b$ in a 2D Euclidean plane, following a homogeneous Poisson point process (HPPP), $\Phi_b$. UEs are independently and uniformly distributed. In this network, channel parameters are given by the values used in the NYU mmWave network simulator \cite{mezzavilla15}, obtained from the NYU mmWave channel model~\cite{akdeniz14,samimi15}. The parameters are summarized in Table I, and elaborated in the following subsections.
 
    
    \begin{table}
    \centering
    \caption{List of notations and channel parameters.}
    \small
    \renewcommand{\arraystretch}{1.2}
    \begin{tabular}{r |l }
    \toprule
    \hspace{-10pt}\bf{Notation} &\hspace{-5pt} \textbf{Meaning}: Parameters\\
    \hline 
    \hspace{-10pt} $f$ & Carrier frequency: 28 GHz\\
    \hspace{-10pt} $\Phi_b$ & BS locations following a HPPP with density $\lambda_b$\\
    \hspace{-10pt} $p_L(r)$ & LoS state probability at distance $r$: $p_L=e^{-0.0149r}$\\
    \hspace{-10pt} $x_o,\; x$ & Serving and interfering BSs or their coordinates\\
    \hspace{-10pt} $\ell^j(r)$ & Path loss at distance $r$ in LoS/NLoS state $j\in\{L,N\}$\\
    \hspace{-10pt} $\alpha_j,\; \beta_j$ & Path loss exponent and gain:\\
    \hspace{-10pt} & $\alpha_L=2$, $\alpha_N=2.92$, $\beta_L=10^{-7.2}$, $\beta_N=10^{-6.14}$\\
    \hspace{-10pt} $n_\text{TX},\; n_\text{RX}$ & \# antennas of a BS and a UE\\
    \hspace{-10pt} $G_o,\; G_x$ & Aligned and misaligned gains \\
    \hspace{-10pt} $f_{G_o}, f_{G_x}$ & Aligned and misaligned gain PDFs\\
    \hspace{-10pt} $K$ & \# clusters $\sim \max\{\textsf{Poiss}(1.8),1\}$ \\
    \hspace{-10pt} $L_k$ & \# subpaths in the $k$-th cluster $\sim \textsf{DiscreteUni}[1,10]$ \\
    \hspace{-10pt} $\theta_{kl}^{\text{RX}}$, $\theta_{kl}^{\text{TX}}$ & Angular spread of subpath $l$ in cluster $k$~\cite{akdeniz14}: \\
    \hspace{-10pt} & $\theta_{k}^{(\cdot)}\hspace{-3pt}\sim \textsf{Uni}[0,2\pi]$, $s_a\sim \max\{\textsf{Exp}(0.178),0.0122\},$\\
    \hspace{-10pt} & $\theta_{kl}^{(\cdot)}=\theta_{k}^{(\cdot)} + (-1)^l s_a/2$\\
    \hspace{-10pt} $P_{kl}$ & Power gain of subpath $l$ in cluster $k$~\cite{samimi15}:\\
    \hspace{-10pt} & $U_k \sim \textsf{Uni}[0,1]$, $Z_k \sim \mathcal{N}(0,4^2)$, $V_{kl} \sim \textsf{Uni}[0,0.6]$, \\
        \hspace{-10pt} & $\tau_{kl}=2.8, P_{kl}=\tfrac{P_{kl}^\prime}{\sum P_{kl}^\prime}$, $P_{kl}^\prime = \frac{U_k^{\tau_{kl}-1}10^{-0.1 Z_k+V_{kl}}}{L_k}$
        \\
    \bottomrule
    \end{tabular}
    \label{Table:Notations}
    \end{table}

    \subsection{BS-UE association and distance attenuation}
    \label{distance_attenuation}
	
	We consider two distance attenuation states: line-of-sight (LoS) and non-line-of-sight (NLoS), respectively denoted with super/subscripts $L$ and $N$ hereafter. For a given link distance~$r$, the LoS and NLoS state probabilities are $p_L(r)$ and $p_N(r) = 1-p_L(r)$. Note that we neglect the outage link state induced by severe distance attenuation, incorporated in \cite{akdeniz14,samimi15}. 
	This simplification does not incur loss of generality for our $\SIR$ analysis, since the received signal powers that correspond to outage are typically negligibly small.

    When a link is in $j\in\{L,N\}$ state with distance $r$, transmitted signals passing through this link experience the following path loss attenuation:
    \begin{align}
    \ell^j(r) = \beta_j r^{-\alpha_j}
    \label{Eq:pathloss}
    \end{align}
    where $\alpha_j$ indicates the path loss exponent and $\beta_j$ is the path loss gain at unit distance.

	Each UE associates with a single BS that provides the maximum average received power, i.e., minimum path loss. According to this rule, the association may not always be with the nearest BS, particularly when the nearest BS is NLoS. On the contrary, the association link can be NLoS if the distance is sufficiently short. To distinguish this difference at a typical UE, we consider the entire BSs $\Phi_b$ to be partitioned into LoS BSs $\Phi_L$ and NLoS BSs $\Phi_N$, and specify the association link LoS/NLoS state by using the subscript $i\in\{L,N\}$. The LoS/NLoS state $j$ of an interfering link is independent of the association LoS/NLoS state $i$.

	It is noted that we assume that the UE density is sufficiently large such that at least one UE is associated with each BS. Then, each BS serves only a single UE per unit time. To focus more on $\SIR$ behaviors, we neglect the effect of scheduling for multiple associations, and consider a typical UE that is always served by its associated BS. The typical UE is regarded as being located at the origin, which does not affect its $\SIR$ behaviors thanks to Slyvnyak's theorem \cite{HaenggiSG} under the HPPP modeling the BS locations. At the typical UE, let $x_o$ and $x\in \Phi_b$ respectively indicate the associated and interfering BSs as well as their coordinates.




    \subsection{Aligned and misaligned beamforming gains}
    For a given link distance, a random channel gain is determined by the NYU channel model that has mmWave channel specific parameters \cite{akdeniz14,samimi15} based on the WINNER~II model~\cite{winner2}. In this model, each link comprises $K$ clusters that correspond to macro-level scattering paths. For cluster $k\leq K$, there exist $L_k$ subpaths, as visualized in Figure~\ref{globalview}. 

    Given a set of clusters and subpaths, the channel matrix of each link is represented as:
    \begin{equation}
    \textbf{H}= \sum_{k=1}^{K}\sum_{l=1}^{L_k}g_{kl} \textbf{u}_{\text{RX}}\left(\theta^{\text{RX}}_{kl}\right) \textbf{u}^*_{\text{TX}}\left(\theta^{\text{TX}}_{kl}\right)
    \label{channel_matrix}
    \end{equation}
    where {$g_{kl}$} is the small-scale fading gain of subpath $l$ in cluster $k$, and $\textbf{u}_{\text{RX}}(\cdot)$ and $\textbf{u}_{\text{TX}}(\cdot)$ indicate the 3D spatial signature vectors of the receiver and transmitter, respectively. Moreover, $\theta^{\text{RX}}_{kl}$ is the angular spread of horizontal angles of arrival (AoA) and  $\theta^{\text{TX}}_{kl}$ is the angular spread of horizontal angles of departure (AoD), both for subpath $l$ in cluster $k$~\cite{akdeniz14}. Note that we consider a planar network and channel, i.e., we neglect vertical signatures by setting their angles to $\pi/2$, as was done in the original channel modeling \cite{mezzavilla15}.

    Next, small-scale fading gain $g_{kl}$ is given as follows:
    \begin{align}
    g_{kl}=\sqrt{P_{kl}}e^{-j2\pi \tau _{kl}f}.
    \label{scale_fading}
    \end{align}
    The term  $P_{kl}$ denotes the power gain of subpath $l$ in cluster $k$, $\tau _{kl}$ is the delay spread induced by different subpath distances, and $f$ indicates carrier frequency. Specific parameters are provided in Table I.


    
    Consider a directional beamforming where the main lobe center of a BS's transmit beam points at its associated UE, while the main lobe center of a UE's receive beam aims at the serving BS. We assume that both beams can be steered in any directions. Therefore, we can generate a beamforming vector for any possible angle in $\[0,2\pi\]$. At a typical UE, the aligned gain $G_o$ is its beamforming gain towards the serving BS at $x_o$. With a slight abuse of notation for the subscript $x_o$, $G_o$ is represented as:
    \begin{align}
    &G_o = |\textbf{w}^T_{\text{RX}_{x_o}} \textbf{H}_{x_o} \textbf{w}_{\text{TX}_{x_o}} |^2\\
    &= \left|  \sum_{k=1}^{K}\sum_{l=1}^{L_k}g_{kl} \left( \textbf{w}^T_{\text{RX}_{x_o}}\textbf{u}_{\text{RX}_{x_o}} \right) \left( \textbf{u}^*_{\text{TX}_{x_o}}\textbf{w}_{\text{TX}_{x_o}} \right) \right|^2
    \label{bf_gain_1}
    \end{align}
    where $\textbf{w}_{\text{TX}_{x_o}} \in \mathbb{C}^{n_{\text{TX}}}$ is the transmitter beamforming vector and $\textbf{w}_{\text{RX}_{x_o}} \in \mathbb{C}^{n_{\text{RX}}}$ is the receiver beamforming vector. Their values are computed as in~\cite{tse_book}. 

    Similarly, a typical UE's misaligned gain $G_x$ is its beamforming gain with an interfering BS at $x$:
    \begin{align}
    &G_{x} = |\textbf{w}^T_{\text{RX}_{x}} \textbf{H}_{x} \textbf{w}_{\text{TX}_{x}} |^2
    \label{bf_gain_2}
    \end{align}
    where $\textbf{w}_{\text{TX}_{x}}$ and $\textbf{w}_{\text{RX}_{x}}$ respectively imply transmitter and receiver beamforming vectors. It is noted that both $G_o$ and $G_x$ incorporate the effects not only of the main lobes but also of the side lobes. 

 	Combining~\eqref{Eq:pathloss},~\eqref{bf_gain_1}, and~\eqref{bf_gain_2}, we can represent $\SIR_i$ as the received $\SIR$ at a typical UE associated with $x_o\in\Phi_i$, given by:
    \begin{align}
    \SIR_i &= \frac{ G_o \ell_i(r_{x_o}^i) }{\sum\limits_{x\in \Phi_L } G_{x} \ell_L(r_x^L) + \sum\limits_{x\in \Phi_N } G_{x} \ell_N(r_x^N)}.
    \label{equation_sinr}
    \end{align}

    For simplicity, we set the BS transmit power to unity, and consider dense BS deployments such that noise is negligibly small compared to interference.


    




    \section{Aligned and Misaligned Gain Distributions}
    \label{Gain_distribution}
    In this section we focus on the beamforming gains $G_o$ in~\eqref{bf_gain_1} and $G_{x}$ in~\eqref{bf_gain_2}, and derive their distributions via curve fitting, a process which is quite relevant for the purpose of our analysis.
    
    On the one hand, the aligned gain $G_o$ is obtained for the desired received signal when the AoA and AoD of the beamforming vectors $\textbf{w}_{\text{TX}_{x_o}}$ and $\textbf{w}_{\text{RX}_{x_o}}$ are aligned with the angles of the spatial signatures $\textbf{u}_{\text{TX}_{x_o}}$ and $\textbf{u}_{\text{RX}_{x_o}}$ in the channel matrix $\textbf{H}_{x_o}$.
    On the other hand, the misaligned gain $G_x$ is calculated for each interfering link with the beamforming vectors and spatial signatures that are not aligned\footnote{Note in fact that beamforming tries to make sure the intended receiver is aligned in the direction from which it is receiving the signal from its serving BS, whereas any interfering BS would be sending to its own associated UE. Hence, the interference between a BS and a UE has a random orientation uniformly distributed in the interval $\left[0,2\pi\right]$.
    For this reason, in~\eqref{bf_gain_2}, the angles of the beamforming vectors $\textbf{w}_{\text{TX}_x}$ and $\textbf{w}_{\text{RX}_x}$, and the angles of the spatial signatures $\textbf{u}_{\text{TX}}$ and $\textbf{u}_{\text{RX}}$ are not aligned as in $G_o$.
    Indeed, they are all i.i.d. uniformly distributed.}.
    Figure~\ref{globalview} shows an example of misalignment between the beam of the desired signal (green beam) and the interfering BSs beams (red beams).    

    In the following subsections, we precisely describe the behaviors of the two gains by extracting their distributions.

    \subsection{Aligned gain distribution}
    \label{alignedt_gain_distribution}
    Using a large number of independent runs of the NYU simulator we empirically evaluated the distribution of the aligned gain $G_o$.
    From the obtained data samples we have noticed that the squared norm of $G_o$ is roughly exponentially distributed $G_o \sim \mathsf{Exp}(\mu_o)$.
    Indeed, the signal's real and imaginary parts are approximately distributed as independent Gaussian random variables.
    This exponential behavior finds an explanation in the small-scale fading effect implemented in the channel model using the power gain term $P_{kl}$ computed as reported in Table~\ref{Table:Notations}.
        
\begin{figure}[t!]
    \centering
    \includegraphics[width=1\columnwidth]{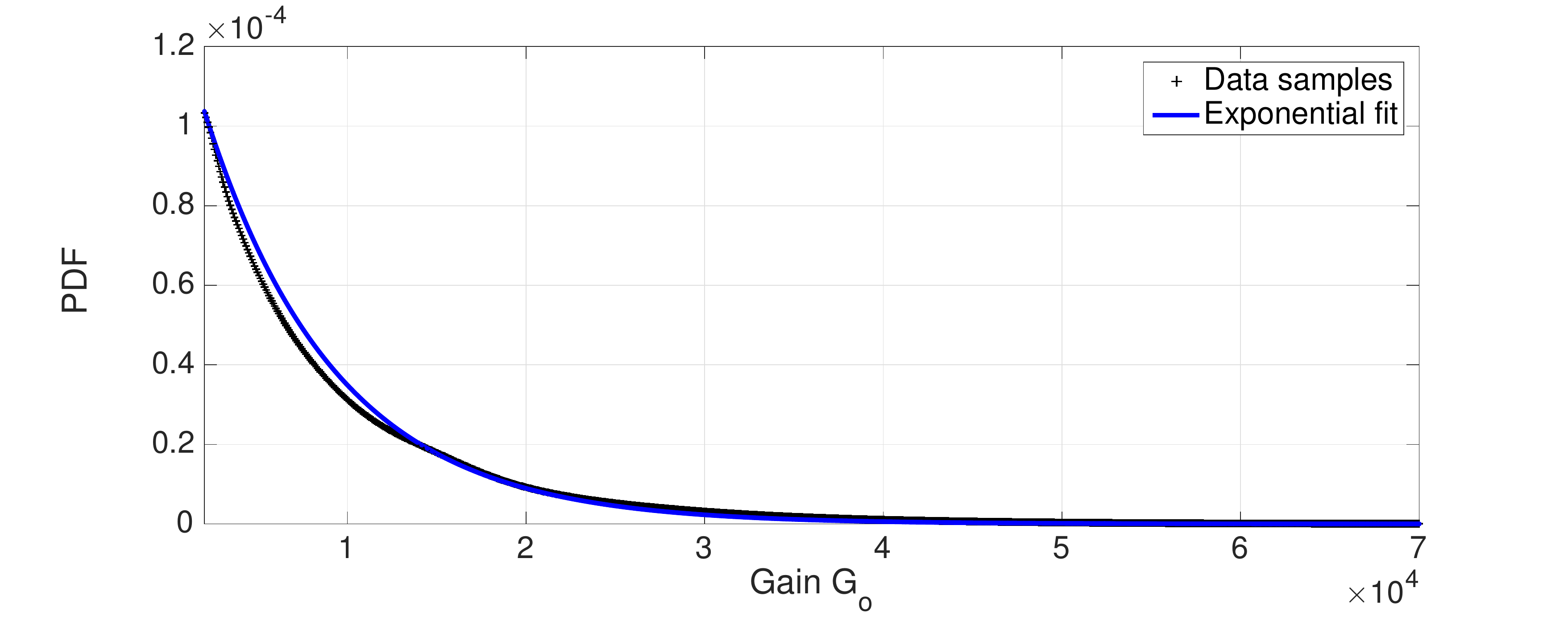}
    \caption{Exponential fit of the PDF of $G_o$, where $f_{G_o}(x) = \mu_o \exp \left(-\mu_o x\right)$. In this case $n_{\text{RX}} = 64$ and $n_{\text{TX}} = 256$.}
    \label{exponential_fit_G0}
    \end{figure}

    We report in Figure~\ref{exponential_fit_G0} an example of the exponential fit of the simulated distribution. The fit has been obtained using the \emph{curve fitting toolbox} of MATLAB.
    
\begin{figure}
    \centering
    \includegraphics[width=1\columnwidth]{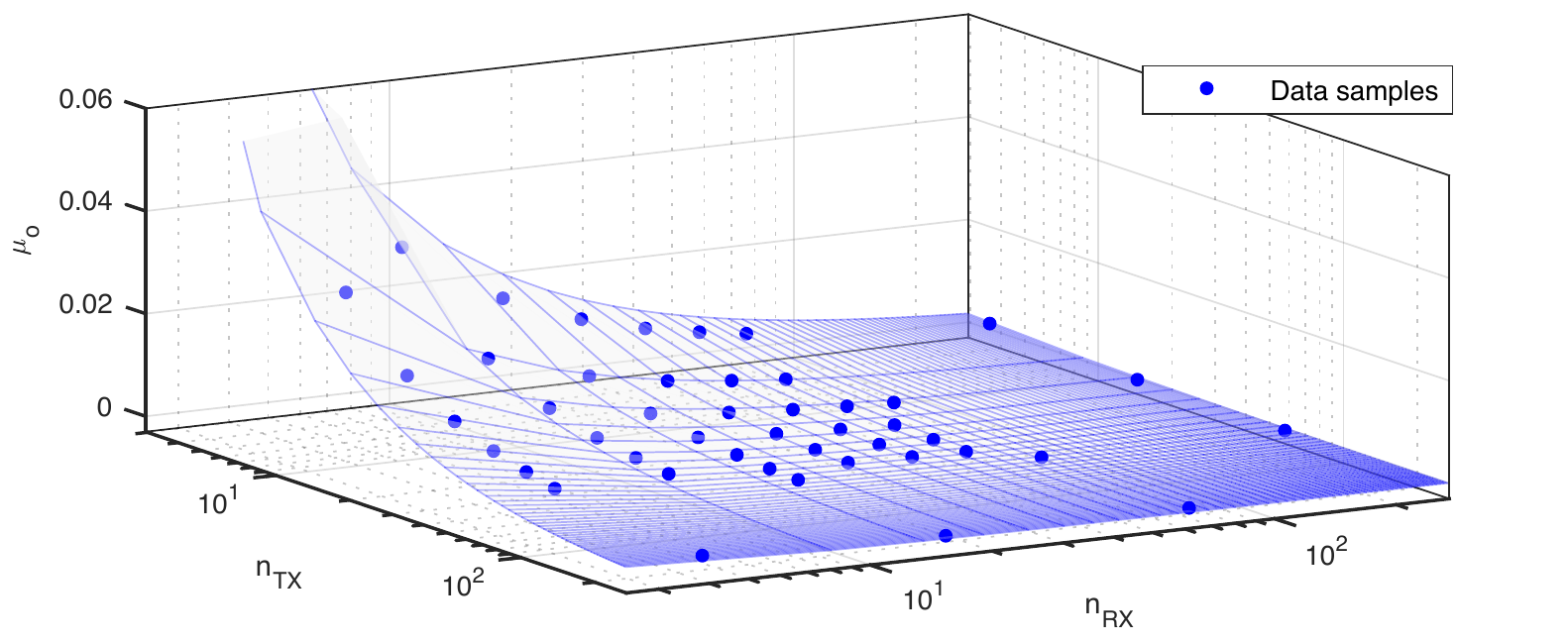}
    \caption{Fit of the $\mu_o$ term varying the number of antenna elements $n_{\text{TX}}$ and $n_{\text{RX}}$. In this case, the fit follows the expression: $\mu_o = 0.814 (n_{\emph{\text{TX}}}n_{\emph{\text{RX}}})^{-0.927}$.}
    \label{power_fit}
    \end{figure}

    For the purpose of deriving a closed-form expression, it is also interesting to evaluate the behavior of $\mu_o$ as a function of the number of antenna elements at both receiver and transmitter sides. 
    For this reason, in our analysis we consider the term $\mu_o$ as a function of the number of antenna elements.
    We show in Figure~\ref{power_fit} the trend of the parameter $\mu_o$ versus the number of antenna elements at the transmitter side $n_{\text{TX}}$ and at the receiver side $n_{\text{RX}}$.
    Again, using the curve fitting toolbox, we have obtained a two dimensional power fit where the value of $\mu_o$ can be achieved as in the following remark.

\begin{remark} At a typical UE, the aligned gain $G_o$ follows an exponential distribution with PDF:
\begin{align}
f_{G_o}(y) = \mu_o e^{-\mu_o y}
\end{align}
where $\mu_o = 0.814 (n_{\emph{\text{TX}}}n_{\emph{\text{RX}}})^{-0.927}$.
\end{remark}
This result provides a fast tool for future calculations.
    Indeed, the expression found for the gain permits to avoid running a detailed simulation every time.
    We note that from a mathematical point of view the surface of the term $\mu_o (n_{\text{TX}},n_{\text{RX}})$ is symmetric.
    In fact, the gain does not depend individually on the number of antennas at the transmitter or receiver sides, but rather on their product, so we can trade the complexity at the BS for that at the UE.

    \begin{figure}[t!]
        \centering
        \begin{subfigure}[b]{0.45\columnwidth}
            \includegraphics[width=\textwidth]{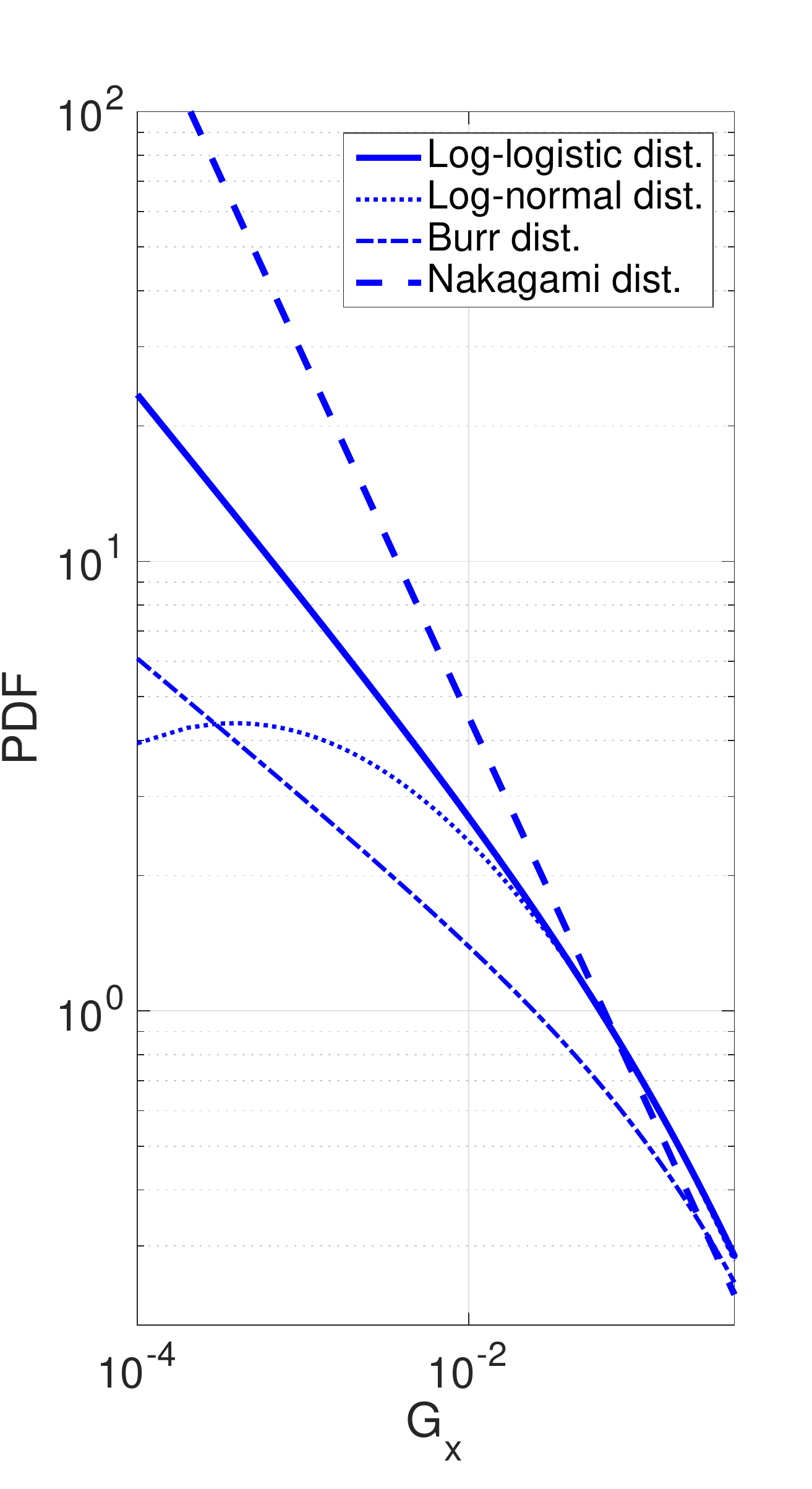}
            \caption{Values close to zero.}
            \label{different_distribution_zero}
        \end{subfigure}
        ~ 
        \begin{subfigure}[b]{0.45\columnwidth}
            \includegraphics[width=\textwidth]{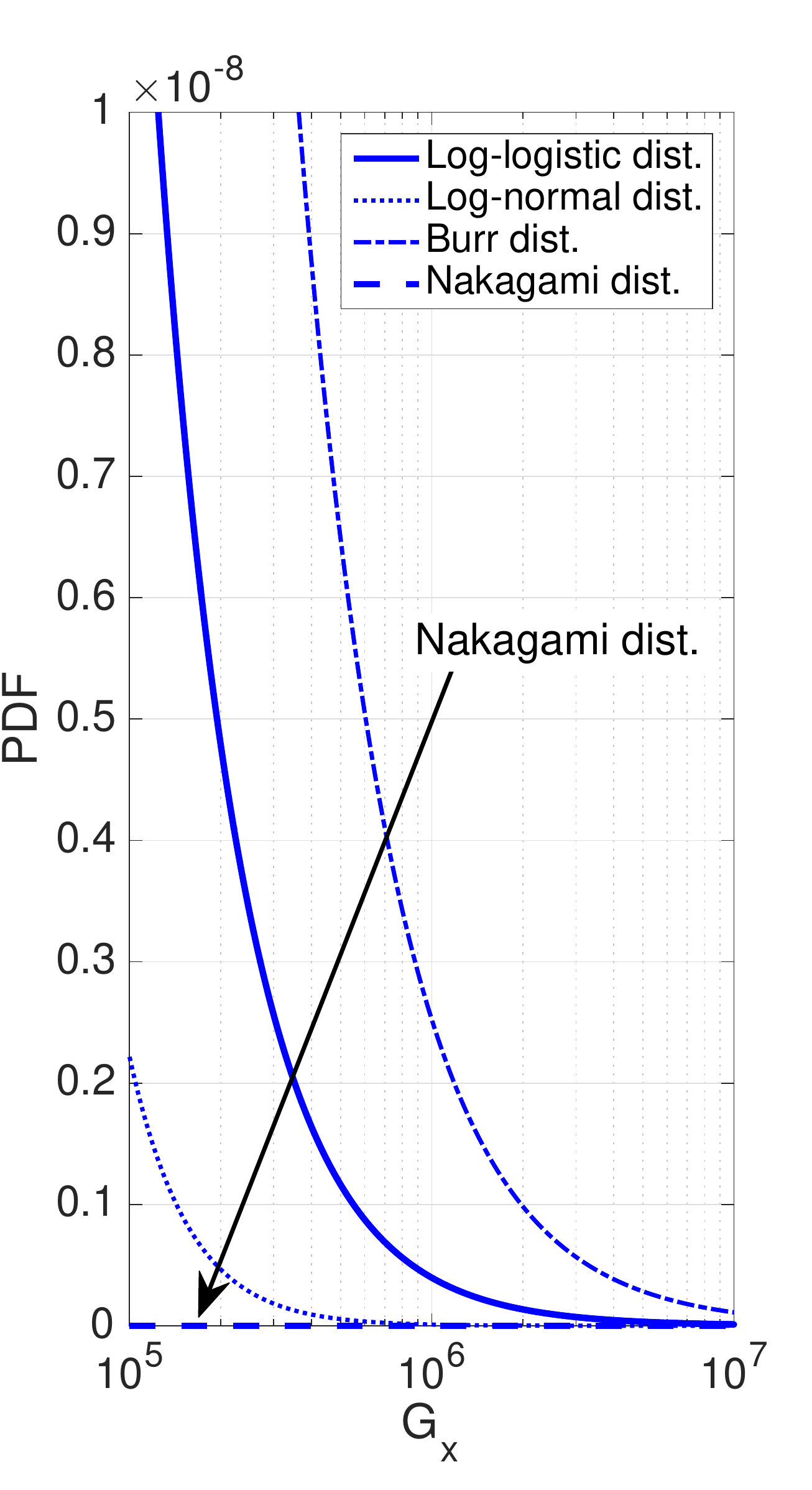}
            \caption{Values in the tail.}
            \label{different_distribution_tail}
        \end{subfigure}
        \caption{Comparison of different distributions for values close to the origin and on the tail. Curves for the case with $n_{\text{TX}} = 256$ and $n_{\text{RX}} = 64$. Nakagami distribution is seen to be the worst case in the set of studied distributions, as it overlaps with the $x$ axis for large values of $G_x$.}
        \label{different_distribution}
    \end{figure}

\setcounter{equation}{17}
\begin{figure*}[b]
    \rule[0.5ex]{\linewidth}{1pt}
    \begin{align}
    \mathcal{L}_{I^j_i} &=  \exp \left( -2 \pi \lambda_b \int_0^{\infty} \left( \int_{\left(\frac{\beta_j r^{\alpha_i}}{\beta_i} \right)^{\frac{1}{\alpha_j}}}^{\infty} \left[1- \exp\left(-   \frac{\ell^j(v)\mu_o T g }{\ell^i(r)} \right)\right] v p_j(v) \mathrm{d}v \right)  f_{G_x}(g) \mathrm{d}g \right)
    \end{align}

    \end{figure*}

    \subsection{Beam misalignment gain distribution}
    \label{beam_misalignment_gain_distribution}

    We also used a large number of independent runs of the NYU simulator in order to extract the distribution of the misaligned gain $G_x$. By simulation, we found that the $G_x$ PDF has a very steep slope in the vicinity of zero. The reason is that sharp transmit/receive beams reduce the main lobe interfering probability, while the dominant interference comes from the side lobes. In addition, we also found that the PDF of $G_x$ is heavy-tailed.
     This follows from the fact that the sharpened main lobe beam leads to strong interference, with its low interfering probability.   
    Moreover, using an accurate beamforming pattern, our study considers also side lobes that are extremely relevant when evaluating the interference perceived from a random direction.
    Simplified models presented in the literature used piece-wise constant patterns that are less accurate than the one used in this study.

    In this respect, we examined candidate distributions satisfying the aforementioned two characteristics, and conclude that the \emph{log-logistic} distribution provides the most accurate fitting result with the simulated misaligned gain as described below.
    \begin{remark} At a typical UE, the misaligned gain $G_x$ is approximated as a log-logistic distribution with PDF:
    \setcounter{equation}{8}\begin{align}
     f_{G_x}(y;a,b) =\frac{\(\frac{b}{a}\)\(\frac{y}{a}\)^{b -1}}{\(1+\(\frac{y}{a}\)^b\)^2}
    \end{align}
    where $a$ and $b$ for different $n_{\emph{\text{TX}}}$ and $n_{\emph{\text{RX}}}$ are provided in Table~\ref{evolution_a_b_factors}.
    \end{remark}    
    It is worth noting that, differently from the exponential fit of $G_0$, generalizing $a$ and $b$ as functions of $n_{\emph{\text{TX}}}$ and $n_{\emph{\text{RX}}}$ is not possible. To overcome this limitation, we report in Table~\ref{evolution_a_b_factors} the values of both parameters for some fixed numbers of antenna elements.

    \begin{table}
    \caption{Log-logistic $G_x$ PDF parameters $(a,b)$ for different $n_{\text{TX}}$ and $n_{\text{RX}}$.}
    \footnotesize
    \centering
    \renewcommand{\arraystretch}{1.1}
    \begin{tabular}{cc||cccc}
    \toprule
    \multicolumn{2}{c||}{\multirow{2}{*}{$(a,b)$}}                                           & \multicolumn{4}{c}{$n_{\text{TX}}$}                                                                                                                                                                                                                                                                                                                                                  \\ \cline{3-6} 
    \multicolumn{2}{c||}{}      & \multicolumn{1}{c|}{4}                                                                           & \multicolumn{1}{c|}{16}                                                                          & \multicolumn{1}{c|}{64}                                                                          & 256                                                                         \\ \hline \hline
    \multicolumn{1}{c|}{\multirow{4}{*}{$n_{\text{RX}}$}} & 4   & \multicolumn{1}{c|}{\begin{tabular}[c]{@{}l@{}}$a = 3.28$\\$b = 0.877$\end{tabular}} & \multicolumn{1}{c|}{\begin{tabular}[c]{@{}l@{}}$a = 2.51$\\ $b = 0.743$\end{tabular}} & \multicolumn{1}{c|}{\begin{tabular}[c]{@{}l@{}}$a = 2.11$\\ $b = 0.722$\end{tabular}} & \begin{tabular}[c]{@{}l@{}}$a = 1.92$\\ $b = 0.709$\end{tabular} \\ \cline{2-6} 
    \multicolumn{1}{c|}{}                                 & 16  & \multicolumn{1}{c|}{\begin{tabular}[c]{@{}l@{}}$a = 2.52$\\ $b = 0.743$\end{tabular}} & \multicolumn{1}{c|}{\begin{tabular}[c]{@{}l@{}}$a = 3.49$\\ $b = 0.656$\end{tabular}} & \multicolumn{1}{c|}{\begin{tabular}[c]{@{}l@{}}$a = 3.28$\\ $b = 0.612$\end{tabular}} & \begin{tabular}[c]{@{}l@{}}$a = 2.89$\\ $b = 0.589$\end{tabular} \\ \cline{2-6} 
    \multicolumn{1}{c|}{}                                 & 64  & \multicolumn{1}{c|}{\begin{tabular}[c]{@{}l@{}}$a = 2.11$\\ $b = 0.722$\end{tabular}} & \multicolumn{1}{c|}{\begin{tabular}[c]{@{}l@{}}$a = 3.28$\\ $b = 0.612$\end{tabular}} & \multicolumn{1}{c|}{\begin{tabular}[c]{@{}l@{}}$a = 2.55$\\ $b = 0.57$\end{tabular}}  & \begin{tabular}[c]{@{}l@{}}$a = 1.98$\\ $b = 0.551$\end{tabular} \\ \cline{2-6} 
    \multicolumn{1}{c|}{}                                 & 256 & \multicolumn{1}{c|}{\begin{tabular}[c]{@{}l@{}}$a = 1.92$\\ $b = 0.709$\end{tabular}} & \multicolumn{1}{c|}{\begin{tabular}[c]{@{}l@{}}$a = 2.89$\\ $b = 0.589$\end{tabular}} & \multicolumn{1}{c|}{\begin{tabular}[c]{@{}l@{}}$a = 1.98$\\ $b = 0.551$\end{tabular}} & \begin{tabular}[c]{@{}l@{}}$a = 1.45$\\ $b = 0.547$
    \end{tabular}
    \\
    \bottomrule
    \end{tabular}
    \label{evolution_a_b_factors}
    \end{table}

    Moreover, we derived upper and lower bound distributions of $G_x$, which correspond respectively to the lower and upper bounds for the $\SIR$ coverage probability in Section V. We found that a \emph{Burr} distribution \cite{burr1942} provides an upper bound for $G_x$, while \emph{log-normal} distribution can be utilized as a lower bound. By simulation, we also found that the heavy-tail behavior of $G_x$ dominates the close-to-zero behavior when deriving the bounds. As Figure 4 illustrates, a Burr distribution overestimates $G_x$, whereas log-normal underestimates the PDF, compared to the log-logistic approximation curve. In spite of the reversed tendency for their close-to-zero behaviors, the bound directions follow from their heavy-tail behaviors. The fitting parameters are specified in the following remark.
    
    \begin{remark} Upper and lower bounds for the misaligned gain $G_x$ can be obtained using Burr and log-normal distributions, respectively with the following PDFs:
 \begin{align}
     \hspace{-10pt} f_{G_x}(y;c,k) &= \tfrac{c k y^{c-1}}{(1+y^c)^{k+1}},  &\begin{cases} c = 0.692 \\ k = 0.518 \end{cases}\\
    \hspace{-10pt}f_{G_x}(y;\sigma,\mu) &= \tfrac{1}{y\sigma \sqrt{2 \pi}}\exp\(-\tfrac{\left(\log y-\mu\right)^2}{2 \sigma^2}\), \hspace{-10pt}&\begin{cases} \sigma = 2.962 \\ \mu = 0.908 \end{cases}
\end{align}
where the specific values reported are for the case with $n_{\text{TX}} = 256$ and $n_{\text{RX}} = 64$. 
\end{remark}

    We additionally tried the curve fitting with a Nakagami-m distribution as follows:
    \begin{align}
    f_{G_x}(y;m,g) = \frac{2m^m}{\Gamma(m) g^m} y^{2m -1} \exp\(-\frac{m}{g}y^2\), _{} \begin{cases} m = 0.099 \\ g = 50.53 \end{cases}
    \end{align}
    where values of $\{m,g\}$ are, as before, for the case with $n_{\text{TX}} = 256$ and $n_{\text{RX}} = 64$. 
    It turned out that a Nakagami distribution underestimates the tail behavior too much, as shown in Figure~\ref{different_distribution}, thereby leading to a loose upper bound for the $\SIR$ coverage probability, as shown in Section~V.

    \section{$\SIR$ Coverage Probability} 
     \label{sinr_coverage}

Consider the $\SIR$ coverage probability $\C(T)$, defined as the probability that the $\SIR$ is no smaller than a target $\SIR$ threshold $T>0$, i.e., $\C(T):=\Pr(\SIR\geq T)$. Based on the aligned and misaligned gain distributions in Section III, in this section we derive $\C(T)$ at a typical UE. 

Thowards this end, we first define the association distance $r_{x_o}^i$ of a typical UE associating with $x_o\in\Phi_i$. By using the law of total probability, $\C$ at a typical UE is represented as follows:
    
    \vspace{-10pt}\small\begin{align}
    \hspace{-0pt}&\C (T) =  \Pr \big( \underbrace{\SIR \geq T,  x_o\in \Phi_L}_{\SIR_L\geq T} \big) + \Pr\big( \underbrace{\SIR \geq T,  x_o\in \Phi_N}_{\SIR_N \geq T} \big)\label{sinr_coverage_definition_pre}\\
    \hspace{-0pt}&= \E_{r_{x_0}^L}\left[\Pr \big( \SIR_L\geq T | r_{x_o}^L \big)\right] + \E_{r_{x_o}^N}\[\Pr\big( \SIR_N \geq T |r_{x_o}^N \big)\].
    \label{sinr_coverage_definition}
    \end{align}\normalsize
According to \cite{Gupta2016}, the PDF of the typical UE's association distance $r_{x_o}^i$ is given as:

    \vspace{-10pt}\small  \begin{align}
    \hspace{-5pt}&f_{r_{x_o}^i}(r) :=f_{|x_o|,i}(r, x_o\in\Phi_i)\\
    \hspace{-5pt}&= 2 \pi \lambda_i(r) r\exp{\Bigg(\hspace{-3pt} - 2    \pi \lambda_b \Bigg[\int_0^{r} \hspace{-3pt}v p_i(v)\mathrm{d}v + \int_0^{ \left(\frac{\beta_{i^\prime} r^{\alpha_i}}{\beta_{i}}\right)^{\frac{1}{\alpha_{i^\prime}}} } \hspace{-15pt}v p_{i'}(v)\mathrm{d}v\Bigg] \Bigg)}
    \end{align}\normalsize
    where $\lambda_i(r)=\lambda_b p_i(r)$ and $i^\prime$ indicates the opposite LoS/NLoS state with respect to $i$.

By exploiting $f_{r_{x_o}^i}(r)$ while applying Campbell's theorem \cite{HaenggiSG} and the $G_o$ distribution in Remark 1, $\C(T)$ is derived as shown in the following proposition.

\begin{proposition} At a typical UE, the $\SIR$ coverage probability $\C(T)$ for a target $\SIR$ threshold $T>0$ is given as:

\vspace{-10pt}\small\begin{align}
\C(T) = \sum_{i\in \{L,N\}} \int_0^{\infty} f_{r_{x_o^i}} \left(r\right) \mathcal{L}_{I_i^L}\left( \frac{\mu_o T}{ \ell^i(r) }\right) \mathcal{L}_{I_i^N}\left(\frac{\mu_o T}{\ell^i(r) }\right) \mathrm{d}r
\end{align}\normalsize
where $\mathcal{L}_{I_i^j}(r)$ for $j\in\{L,N\}$ is given at the bottom of this page.
\\\noindent\text{Proof:} \text{\normalfont\ See Appendix.} \hfill $\blacksquare$
\end{proposition}


    Note that $1/\mu_o$ is the mean aligned gain in Remark 1. The misaligned gain PDF $f_{G_x}(y)$ and its corresponding parameters are provided in Remarks 2 and 3 as well as in Table \ref{evolution_a_b_factors}. The term $p_i$ is the LoS/NLoS channel state probability defined in Section~\ref{distance_attenuation}.

    \section{Numerical Results}
    \label{numerical_results}
    In this section, we provide some results numerically computed using the analytical expressions derived in the previous sections. 
    We report plots of the $\SIR$~\emph{coverage} probability, for a network that operates at 28~GHz and has a BS density $\lambda_b$ equal to 100 BSs per km$^2$. 

\begin{figure}
    \centering
    \includegraphics[width=0.95\columnwidth]{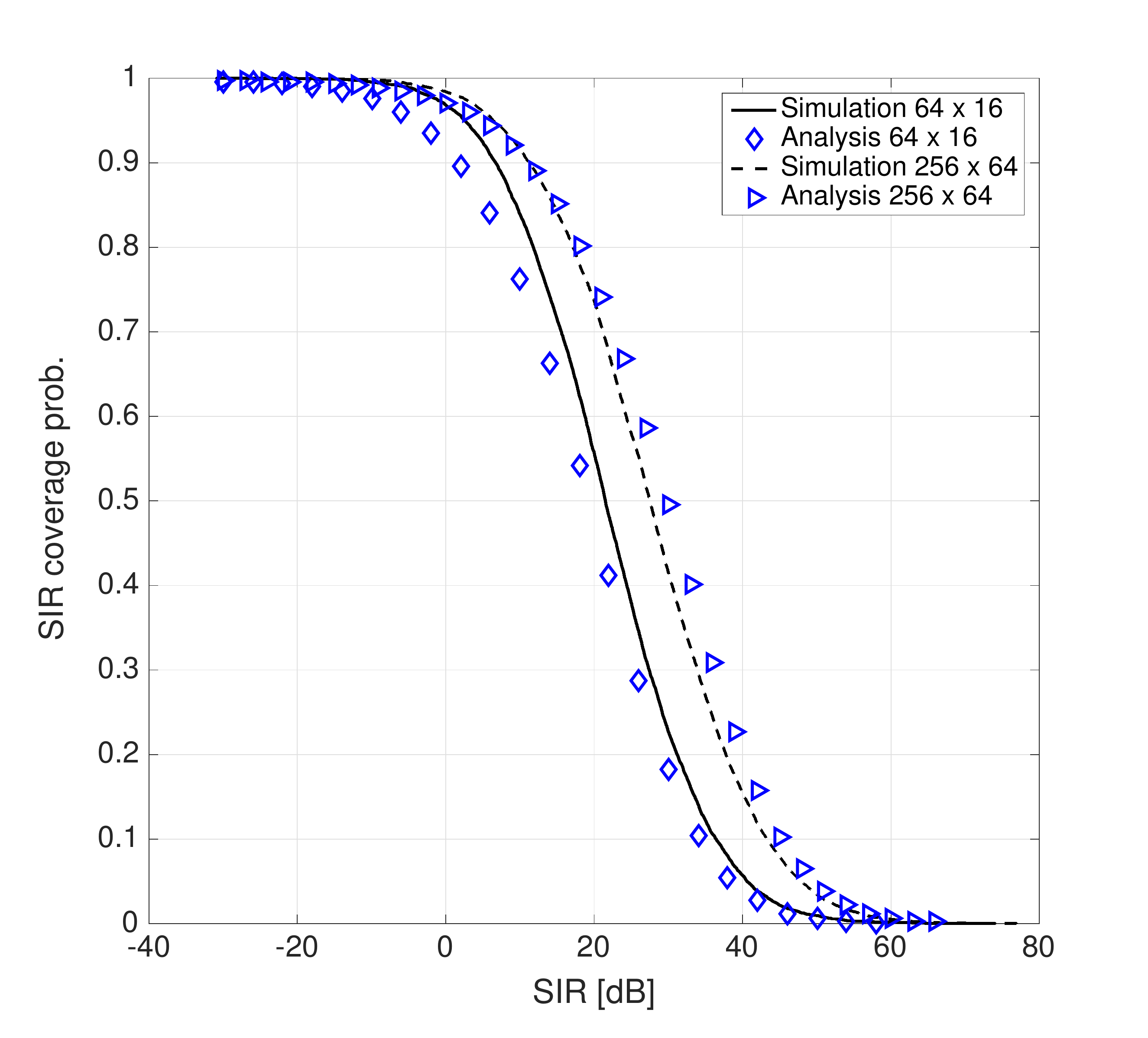}
    \caption{$\SIR$ coverage probability with the \emph{log-logistic} misaligned gain PDF for $\{n_{\text{TX}},n_{\text{RX}}\}=\{64, 16\}$ and $\{256, 64\}$.}
    \label{sir_cdf}
    \end{figure}

    Figure~\ref{sir_cdf} compares the coverage probability for two different antenna configurations and validates our analysis with simulations made using the NYU mmWave channel model.
    We have plotted curves of the $\SIR$ coverage probability for configurations with 256$\times$64 and 64$\times$16 antennas, and in both cases we can observe that the numerical results closely follow the simulation.
    It is important to highlight that, in order to properly compare the simulation with the numerically computed analysis, our simulations have considered the NYU simulator completely with LoS and NLoS states and with a simulation area large enough to consider all the significant received signals, thus avoiding edge effects.
    Furthermore, the figure shows that a larger number of antenna elements results in a higher $\SIR$ due to the higher gain achieved by beamforming as expected. 

\begin{figure}
    \centering
    \includegraphics[width=0.95\columnwidth]{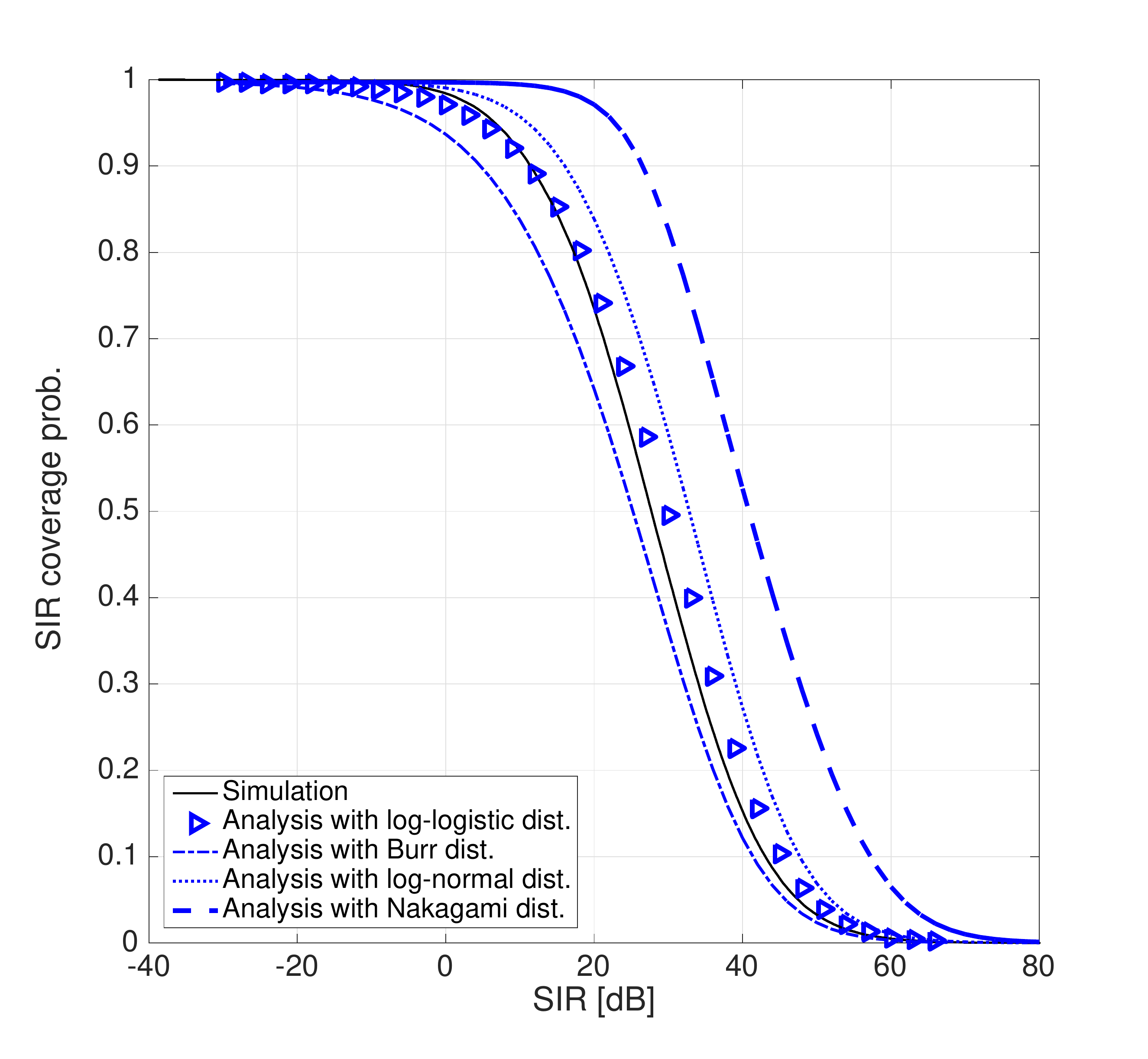}
    \caption{$\SIR$ coverage probability with different misaligned gain PDFs: \emph{log-logistic} (approx.), \emph{Burr} (lower bound), \emph{log-normal} (upper bound), and \emph{Nakagami-m} (loose upper bound) distributions for $\{n_{\text{TX}},n_{\text{RX}}\}=\{256, 64\}$.}
    \label{bounds}
    \end{figure}

    Figure~\ref{bounds} shows the $\SIR$ coverage probability curve using different distributions when modeling the misaligned gain in the interfering links. 
    The plot shows the upper and lower bounds that can be achieved using log-normal and Burr distributions, respectively. The figure also highlights the large gap between the real simulated curve and the one obtained when the misaligned gain is modeled by Nakagami distribution.
    
    We are not reporting the results of a direct comparisons between our analysis and preceding analyses because, due to the differences in the channel model and in the beam pattern, it is extremely complicated to find an exact correspondence between parameters of different models for a fair comparsion. 

    \section{Conclusions and Future Directions}
    \label{conclusion_and_future_works}
    In this paper we discussed how to marry a realistic mmWave channel model and its tractable large-scale network analysis, by applying the NYU mmWave channel model to stochastic geometric $\SIR$ coverage analysis.
    For this purpose, we first defined aligned and misaligned gains, containing the aggregate channel behaviors respectively of the desired and interfering links.
    By using a curve fitting tool, we have identified an exponential behavior for the aligned gain.
    We have also determined a log-logistic behavior for the misaligned gain, as well as Burr and log-normal comportments for its upper and lower bounds, respectively.     
    We applied these fitted distributions to the $\SIR$ coverage probability derived by using stochastic geometry.
    This enables a realistic-yet-tractable $\SIR$ analysis in mmWave cellular networks, thereby avoiding the need for time-consuming Monte Carlo simulations.

    The proposed way of semi-simulated large-scale network analysis is versatile. With this framework, it could be interesting to investigate other 5G scenarios including different mmWave carrier frequencies and antenna configurations.
    
\section*{Appendix -- Proof of Proposition 1}
\setcounter{equation}{18}

Consider  the joint probability $\Pr \big( \SIR \geq T,  x_o\in \Phi_i\big)$ in~\eqref{sinr_coverage_definition_pre} when a typical UE associates with a BS in state $i\in \{L,N\}$. According to~\eqref{sinr_coverage_definition}, it is represented as follows:

\vspace{-10pt}\small\begin{align}
\Pr &\big( \SIR \geq T,  x_o\in \Phi_i\big) = \E_{r_{x_0}^i}\left[\Pr \big( \SIR_i \geq T | r_{x_o}^i \big)\right]\\
&=\int_0^{\infty} f_{r_{x_o}^i} \left(r\right)  \Pr \left( G_o \geq \frac{T (I_i^L + I_i^N)}{ \ell^i(r)} \Big | r \right) \mathrm{d}r\\
&= \int_0^{\infty} f_{r_{x_o}^i} \left(r\right)  \mathcal{L}_{I_i^L}\left( \frac{\mu_o T}{ \ell^i(r)} \right) \mathcal{L}_{I_i^N}\left(\frac{\mu_o T}{\ell^i(r)}\right) \mathrm{d}r. \label{Eq:SIR_PfProp1}
\end{align}\normalsize
The last step results from the exponentially distributed $G_o$ as specified in Remark 1, using the Laplace functional $\mathcal{L}_X(s):=\E_X[e^{s X}]$.

	Given the typical UE's association with $x_o\in\Phi_i$, the Laplace functional of the interference from the BSs in $\Phi_j$ for $j\in\{L,N\}$  is obtained as follows:	
	
	\vspace{-10pt}
    \footnotesize
    \begin{align}
    \hspace{-7pt}&\mathcal{L}_{I_i^j}(s) =  \E_{\Phi_{j},G_x}\left[\exp{\left(-s \sum_{x \in \Phi_j } G_x  \ell^j(r_x) \right)}\right]\\
    \hspace{-7pt}&\overset{(a)}{=} \E_{\Phi_j}\left[\prod_{x \in \Phi_j} \E_{G_x}\left[ \exp{\left(-s G_x  \ell^j(r_x) \right)}\right]\right] \\
    \hspace{-7pt}&\overset{(b)}{=} \exp{\left( -2 \pi \lambda_b \int_{\left( \frac{\beta_j r^{\alpha_i}}{\beta_i} \right)^\frac{1}{\alpha_j}}^{\infty} \left(1- \E_{G_x} \left[e^{-sG_x  \ell^j(v)}\right]\right) v p_j(v) \mathrm{d}v \right)}   \\
    \hspace{-7pt}&= \exp \Bigg( -2 \pi \lambda_b \E_{G_x} \left[ \int_{\left( \frac{\beta_j r^{\alpha_i}}{\beta_i} \right)^\frac{1}{\alpha_j}}^{\infty} \left(1-  e^{-   \frac{\mu_o T G_x \ell^j(v)}{\ell^i(r)} }\right) v p_j(v) \mathrm{d}v \right] \Bigg)
    \end{align}\normalsize
    where $(a)$ follows from i.i.d. $G_x$'s and $(b)$ from applying the probability generating functional (PGFL) of a HPPP \cite{HaenggiSG}. The last step comes from the independence between interferer locations and $G_x$'s. Combining this result with~\eqref{Eq:SIR_PfProp1} and~\eqref{sinr_coverage_definition} and applying the law of total probability completes the proof. $\hfill\blacksquare$



    \bibliographystyle{IEEEtran}
    \bibliography{biblio}

\begin{thebibliography}{10}
\providecommand{\url}[1]{#1}
\csname url@samestyle\endcsname
\providecommand{\newblock}{\relax}
\providecommand{\bibinfo}[2]{#2}
\providecommand{\BIBentrySTDinterwordspacing}{\spaceskip=0pt\relax}
\providecommand{\BIBentryALTinterwordstretchfactor}{4}
\providecommand{\BIBentryALTinterwordspacing}{\spaceskip=\fontdimen2\font plus
\BIBentryALTinterwordstretchfactor\fontdimen3\font minus
  \fontdimen4\font\relax}
\providecommand{\BIBforeignlanguage}[2]{{%
\expandafter\ifx\csname l@#1\endcsname\relax
\typeout{** WARNING: IEEEtran.bst: No hyphenation pattern has been}%
\typeout{** loaded for the language `#1'. Using the pattern for}%
\typeout{** the default language instead.}%
\else
\language=\csname l@#1\endcsname
\fi
#2}}
\providecommand{\BIBdecl}{\relax}
\BIBdecl

\bibitem{mmwave3gpp}
{3GPP TR 38.803 v2.0.0}, ``{TR} for {S}tudy on {N}ew {R}adio {A}ccess
  {T}echnology: {RF} and co-existence aspects,'' Tech. Rep., 2017.

\bibitem{roh2014}
W.~Roh, J.~Y. Seol, J.~Park, B.~Lee, J.~Lee, Y.~Kim, J.~Cho, K.~Cheun, and
  F.~Aryanfar, ``Millimeter-wave beamforming as an enabling technology for 5{G}
  cellular communications: theoretical feasibility and prototype results,''
  \emph{IEEE Communications Magazine}, vol.~52, no.~2, pp. 106--113, February
  2014.

\bibitem{bai14}
T.~Bai, A.~Alkhateeb, and R.~W. Heath, ``Coverage and capacity of
  millimeter-wave cellular networks,'' \emph{IEEE Communications Magazine},
  vol.~52, no.~9, pp. 70--77, September 2014.

\bibitem{bai15}
T.~Bai and R.~W. Heath, ``Coverage and {R}ate {A}nalysis for
  {M}illimeter-{W}ave {C}ellular {N}etworks,'' \emph{IEEE Transactions on
  Wireless Communications}, vol.~14, no.~2, pp. 1100--1114, Feb 2015.

\bibitem{direnzo2015}
M.~D. Renzo, ``Stochastic {G}eometry {M}odeling and {A}nalysis of
  {M}ulti-{T}ier {M}illimeter {W}ave {C}ellular {N}etworks,'' \emph{IEEE
  Transactions on Wireless Communications}, vol.~14, no.~9, pp. 5038--5057,
  Sept 2015.

\bibitem{park2016}
J.~Park, S.~L. Kim, and J.~Zander, ``Tractable {R}esource {M}anagement {W}ith
  {U}plink {D}ecoupled {M}illimeter-{W}ave {O}verlay in {U}ltra-{D}ense
  {C}ellular {N}etworks,'' \emph{IEEE Transactions on Wireless Communications},
  vol.~15, no.~6, pp. 4362--4379, June 2016.

\bibitem{direnzo2013}
M.~D. Renzo, A.~Guidotti, and G.~E. Corazza, ``Average {R}ate of {D}ownlink
  {H}eterogeneous {C}ellular {N}etworks over {G}eneralized {F}ading {C}hannels:
  {A} {S}tochastic {G}eometry {A}pproach,'' \emph{IEEE Transactions on
  Communications}, vol.~61, no.~7, pp. 3050--3071, July 2013.

\bibitem{akdeniz14}
M.~Akdeniz, Y.~Liu, M.~Samimi, S.~Sun, S.~Rangan, T.~Rappaport, and E.~Erkip,
  ``Millimeter {W}ave {C}hannel {M}odeling and {C}ellular {C}apacity
  {E}valuation,'' \emph{IEEE Journal on Selected Areas in Communications},
  vol.~32, no.~6, pp. 1164--1179, June 2014.

\bibitem{samimi15}
M.~K. Samimi and T.~S. Rappaport, ``3-{D} statistical channel model for
  millimeter-wave outdoor mobile broadband communications,'' in \emph{IEEE
  International Conference on Communications (ICC)}, June 2015, pp. 2430--2436.

\bibitem{mezzavilla15}
\BIBentryALTinterwordspacing
M.~Mezzavilla, S.~Dutta, M.~Zhang, M.~R. Akdeniz, and S.~Rangan, ``5{G}
  {M}m{W}ave {M}odule for the {N}s-3 {N}etwork {S}imulator,'' in
  \emph{Proceedings of the 18th ACM International Conference on Modeling,
  Analysis and Simulation of Wireless and Mobile Systems}, ser. MSWiM
  '15.\hskip 1em plus 0.5em minus 0.4em\relax New York, NY, USA: ACM, 2015, pp.
  283--290. [Online]. Available:
  \url{http://doi.acm.org/10.1145/2811587.2811619}
\BIBentrySTDinterwordspacing

\bibitem{ford16}
\BIBentryALTinterwordspacing
R.~Ford, M.~Zhang, S.~Dutta, M.~Mezzavilla, S.~Rangan, and M.~Zorzi, ``A
  {F}ramework for {E}nd-to-{E}nd {E}valuation of 5{G} mm{W}ave {C}ellular
  {N}etworks in {N}s-3,'' in \emph{Proceedings of the Workshop on Ns-3}, ser.
  WNS3 '16.\hskip 1em plus 0.5em minus 0.4em\relax New York, NY, USA: ACM,
  2016, pp. 85--92. [Online]. Available:
  \url{http://doi.acm.org/10.1145/2915371.2915380}
\BIBentrySTDinterwordspacing

\bibitem{rebato16}
M.~Rebato, M.~Mezzavilla, S.~Rangan, and M.~Zorzi, ``Resource {S}haring in 5{G}
  mm{W}ave {C}ellular {N}etworks,'' in \emph{IEEE Conference on Computer
  Communications Workshops (INFOCOM WKSHPS)}, April 2016, pp. 271--276.

\bibitem{HaenggiSG}
M.~Haenggi, \emph{{Stochastic Geometry for Wireless Networks}}.\hskip 1em plus
  0.5em minus 0.4em\relax Cambridge Univ. Press, 2013.

\bibitem{winner2}
P.~Kyosti and {et~al.}, ``{WINNER} {II} channel model,'' \emph{Technical Report
  IST-WINNER D1.1.2 ver 1.1}, September 2007.

\bibitem{tse_book}
D.~Tse and P.~Viswanath, \emph{Fundamentals of Wireless Communication}.\hskip
  1em plus 0.5em minus 0.4em\relax New York, NY, USA: Cambridge University
  Press, 2005.

\bibitem{burr1942}
\BIBentryALTinterwordspacing
I.~W. Burr, ``Cumulative frequency functions,'' \emph{The Annals of
  Mathematical Statistics}, vol.~13, no.~2, pp. 215--232, 1942. [Online].
  Available: \url{http://www.jstor.org/stable/2235756}
\BIBentrySTDinterwordspacing

\bibitem{Gupta2016}
A.~K. Gupta, J.~G. Andrews, and R.~W. Heath, ``On the {F}easibility of
  {S}haring {S}pectrum {L}icenses in mm{W}ave {C}ellular {S}ystems,''
  \emph{IEEE Transactions on Communications}, vol.~64, no.~9, pp. 3981--3995,
  Sept 2016.

\end{thebibliography}

    \end{document}